# Partially coherent quantum degenerate electron matter waves

**Sam Keramati**[1,*], **Eric Jones**[1], **Jeremy Armstrong**[2], **Herman Batelaan**[1,**]

[1]Department of Physics and Astronomy, University of Nebraska-Lincoln, Lincoln, NE 68588, USA
[2]Department of Physics and Astronomy, University of Nebraska at Kearny, Kearny, NE 68849, USA

---

Authors to whom any correspondence should be addressed:
[*]sam.keramati@huskers.unl.edu
[**]hbatelaan@unl.edu

**ABSTRACT**

The use of electron beams is ubiquitous; electron microscopy, scanning tunneling microscopy, electron lithography, and electron diffractometry all use well-collimated and focused beams. On the other hand, quantum degenerate electron beams do currently not exist. The realization of such beams may impact all electron beam technologies and are interesting to pursue. Past attempts to reach degeneracy were hampered by the low degeneracy of continuously emitting electron sources. With the recent advent of ultra-short electron pulses, high degeneracy is expected. Coulomb repulsion and low quantum coherence are hurdles that need to be overcome. A quantum analysis of the electron degeneracy for partially coherent pulsed electron sources is presented and two-particle coincidence spectra are obtained for source parameters that are currently available. The conclusive demonstration of the fermionic Hanbury Brown-Twiss (HBT) effect for free electrons is shown to be within reach, and our results support the claim that femto-second nanotip electron sources, both polarized and unpolarized, can manifest partial to complete quantum degeneracy with appreciable signal-to-noise-ratios for free electron pulses notwithstanding their small particle numbers.

## I. Introduction

More than a decade ago, the observation of electron degeneracy pressure for a free continuous electron beam was reported [1]. The free electron Hanbury-Brown Twiss (HBT) effect has not been repeated in other experiments and the observed defining feature, that is antibunching, has been explained by Coulomb repulsion rather than by electron quantum degeneracy in later theoretical papers [2,3]. Currently the production of degenerate free electron beams is an open problem. The HBT effect for photons [4] was pivotal in the development of the field of quantum optics, and the observation of the same effect for free electron beams may open up a new field of research. Additionally, ever more dense electron pulses are being produced for Ultrafast Electron Diffraction (UED), Ultrafast Electron Microscopy (UEM), and laser wake field acceleration, where single shot electron diffraction imaging of single molecules is one of the major outstanding challenges [5–7]. The question on what limits the density of single electron pulses, degeneracy pressure or Coulomb interaction, is both relevant for quantum mechanics and for applications, such as making molecular movies.

Shortly after the first experiment on the electron HBT effect [1], pulsed electron sources from nanotips were developed [8,9]. The estimated degeneracy of these sources is high [10]. In the HBT experiment, the nanotip electron sources that produced a continuous emission of electrons, were claimed to provide a world record degeneracy of $10^{-4}$, while the femto-second pulsed nanotip electron sources are estimated to give an electron degeneracy of $10^{-1}$ and the corresponding antibunching signal is predicted to be strong. Moreover, in a 1-D approach it has been shown that the Coulomb interaction and electron degeneracy both contribute to the antibunching HBT effect with equal proportions for the new pulsed nanotip electron sources [10]. Experiments are now under development to observe electron degeneracy with such a pulsed source unambiguously.

A difficulty with the experiments is that the degeneracy will not attain the maximum possible value of one. At such a value a fully coherent description of the quantum mechanical wavefunction would be sufficient. However, at lesser values the electron sources and pulses are partially coherent. To address the question how partial coherence influences the dynamics of two-electron wavepackets and the experimental antibunching signature, a theoretical framework is developed in this paper. We conclude that coincidence techniques can be used to attempt the observation of antibunching for two-particle detection, even when the electron pulse is partially coherent to the extent as expected for current pulsed nanotip electron sources. Our theoretical approach is intended to make estimates of the electron count rate signal and the coincidence signal as a function of time for design parameters of table-top experiments.

Historically, R. Hanbury Brown and R. Q. Twiss (HBT) demonstrated that intensity fluctuations of photons emitted from independent thermal light sources show correlations when two detectors are located close to each other; for detectors outside of a single coherence volume no correlations were observed. The resultant photon bunching, which gives rise to an enhancement in the photon coincidence counts on two detectors within a single coherence volume, can both be justified by classical intensity fluctuation theory as well as quantum two-photon interference theory [11]. The former attributes the enhancement in photon counts to an increase in the normalized second order correlation function, which is defined as

$$g^{(2)} = \frac{\langle I_1 I_2 \rangle}{\langle I_1 \rangle \langle I_2 \rangle}. \quad (1)$$

Here, $I_1$ and $I_2$ are time-dependent intensities detected by each of the two detectors. The brackets indicate averaging over time. The latter approach provides a full quantum optical treatment of the problem based on the bosonic nature of the photons [12]. Identical bosons tend to bunch due to the enhancement in the second order correlation function, while identical fermions tend to antibunch because of the Pauli exclusion principle (PEP) which inhibits simultaneous phase-space occupation of any coherence volume. HBT-type correlations have been used to probe quantum mechanical particle statistics for various kinds of identical particles. Antibunching of thermal neutrons [13], antibunching in a Fermi gas of $^{40}K$ atoms [14], and the HBT effect for $^{3}He$ and $^{4}He$, with the former being a fermion and the latter a boson [15], were reported about a decade ago. The HBT effect with electrons in semiconductor devices typically consisting of electron sources, drains, and mesoscopic beam splitters was observed [16–18]. In such experiments, high phase-space degeneracies, i.e. the number of electrons per phase-space cell volume [10], are available, and Coulomb repulsion between electrons is screened by space charge [3]. For free electron beams a conclusive demonstration of the HBT remains a challenge. A tour-de-force experiment was reported in 2002 [1]. A reduction of $1.26\times10^{-3}$ in the coincidence count rate of free electrons in a continuous beam emitted from a high-brightness cold field-emission tip was demonstrated. However, later analyses revealed that such small signals could have been dominated by Coulombic interactions between the free electrons in the beam casting doubt on the original claim that the direct manifestation of the PEP had been observed [2,3].

## II. Results and Discussion

The HBT effect can be described by the second-order correlation function given in equation (1). However, this is not the only possible function to characterize the two-particle interference [19]. The probability distribution $p(n,T)$ for $n$ counts to be registered by a single detector in a time interval $T$, as well as the joint detection probability density distribution $P(\tau)$ for the time delay $\tau$ to occur between two successive detection events can be appropriate functions. In our analytical model, it is most convenient to obtain the two-electron joint detection probability on a detector located in the far-field of the emitter as a function of the detection time delay which is $P(\tau)$. Non-relativistic quantum mechanics suffices for our demonstration of the enhancement in the HBT effect with pulsed electron beams compared with continuous beams. In addition, we neglect multi-particle interference effects that involve more than two electrons since they are rare for the experimental scenario considered here, namely low particle numbers in each pulse and short coherence times. We consider the wavepacket propagation along the beam axis. This is justified noting to the fact that in the experimental arrangement [1], a quadrupole lens is used to expand the coherence volume appropriately in the transverse plane. This in effect trivializes the physics in the perpendicular directions, conveniently leaving the implications of the PEP pertinent to the longitudinal axis which makes the 1-D treatment valid in the real situation.

The two-electron wavefunction is the tensor product of a spatial part $\varphi(\vec{r}_1,\vec{r}_2)$ and the spinor $\chi(s,m)$, where $\vec{r}_1$ and $\vec{r}_2$ are the position vectors of the two electrons labeled with $1$ and $2$, $s$ gives the total spin eigenvalue, and $m$ gives the eigenvalue of the spin operator in the $z$-direction. The symmetric triplet states are denoted by $\chi_S\left(s=1,m=1,0,-1\right)$, and the antisymmetric singlet state by $\chi_{AS}\left(s=0,m=0\right)$. According to the spin statistics theorem, the total wavefunction of the two electrons

$$\psi\left(x_1,x_2;s,m\right)=\varphi\left(\vec{r}_1,\vec{r}_2\right)\chi\left(s,m\right), \tag{2}$$

must be antisymmetric. Consequently, an antisymmetric spatial wavefunction $\varphi_{AS}\left(\vec{r}_1,\vec{r}_2\right)$ should be used with $\chi_S$ and a symmetric one $\varphi_S\left(\vec{r}_1,\vec{r}_2\right)$ with $\chi_{AS}$. The joint detection probability of two electrons is then given by

$$P_{S,AS}\left(\vec{r}_1,\vec{r}_2\right)=\left|\varphi_{S,AS}\left(\vec{r}_1,\vec{r}_2\right)\right|^2, \tag{3}$$

where the normalized symmetric and antisymmetric spatial wavefunctions are

$$\varphi_{S,AS}\left(\vec{r}_1,\vec{r}_2\right)=\frac{1}{\sqrt{2}}\left[\varphi_1\left(\vec{r}_1\right)\varphi_2\left(\vec{r}_2\right)\pm\varphi_1\left(\vec{r}_2\right)\varphi_2\left(\vec{r}_1\right)\right], \tag{4}$$

with the (minus) plus sign for the (anti-)symmetric function. The subscripts in the single-electron wavefunctions $\varphi_i$ with $i=1,2$ denote the particle exchange number. Substituting equation (4) into equation (3) one obtains

$$P_{S,AS}\left(\vec{r}_1,\vec{r}_2\right)=\frac{1}{2}\left\{\left|\varphi_1\left(\vec{r}_1\right)\right|^2\left|\varphi_2\left(\vec{r}_2\right)\right|^2+\left|\varphi_1\left(\vec{r}_2\right)\right|^2\left|\varphi_2\left(\vec{r}_1\right)\right|^2\pm2\,\mathrm{Re}\left[\varphi^*{}_1\left(\vec{r}_1\right)\varphi^*{}_2\left(\vec{r}_2\right)\varphi_1\left(\vec{r}_2\right)\varphi_2\left(\vec{r}_1\right)\right]\right\}, \tag{5}$$

for the joint detection probability density of an electron pair. It is the two-electron interference term given by the overlap of the single-particle wavefunctions in $\varphi_{AS}$ that gives rise to the HBT antibunching effect. For an electron source which coherently emits singlet pairs, $\chi_{AS}$ (associated with $\varphi_S$), constructive interference will resultantly occur that resembles boson bunching. This is not a violation of PEP since such electron pairs with antiparallel spins do not occupy the same phase-space cell. A fully spin-polarized and fully coherent electron source emitting two spin-up electrons at a time, with $\chi_S\left(s=1,m=1\right)$ associated with $\varphi_{AS}$, would thus be expected to give rise to an HBT antibunching effect with perfect contrast, i.e. with zero joint detection probability density for perfect coincidence of the electron pairs. The question is, how will the partial coherence of a more realistic electron source affect the measurable HBT contrast of the joint detection probability density. To this end, we need to introduce partial quantum coherence to our analysis.

Consider a pair of electrons with a coherence time $T_c$ carried by a pulse with duration $T_{pulse} \geq T_c$. The limit $T_{pulse} \to \infty$ will asymptotically reach the continuous beam condition that was experimentally studied in the past [1]. Figure 1(a) shows the discretization of the time axis into equal time intervals. Each coherence time interval $T_c$ is split into two equal intervals each with duration $a = T_c/2$. Every such interval then corresponds to a phase-space cell that can bear up to one electron according to the PEP. This way a pair of electrons in any two neighboring intervals are coherent, namely quantum mechanically correlated, and incoherent, namely uncorrelated, otherwise. As pointed out in [20], *"it is possible to regard a "realistic" beam of particles either as a mixture of plane-wave states (monoenergetic free particle states) or as a mixture of wave-packet states."* The second approach is what we implement here.

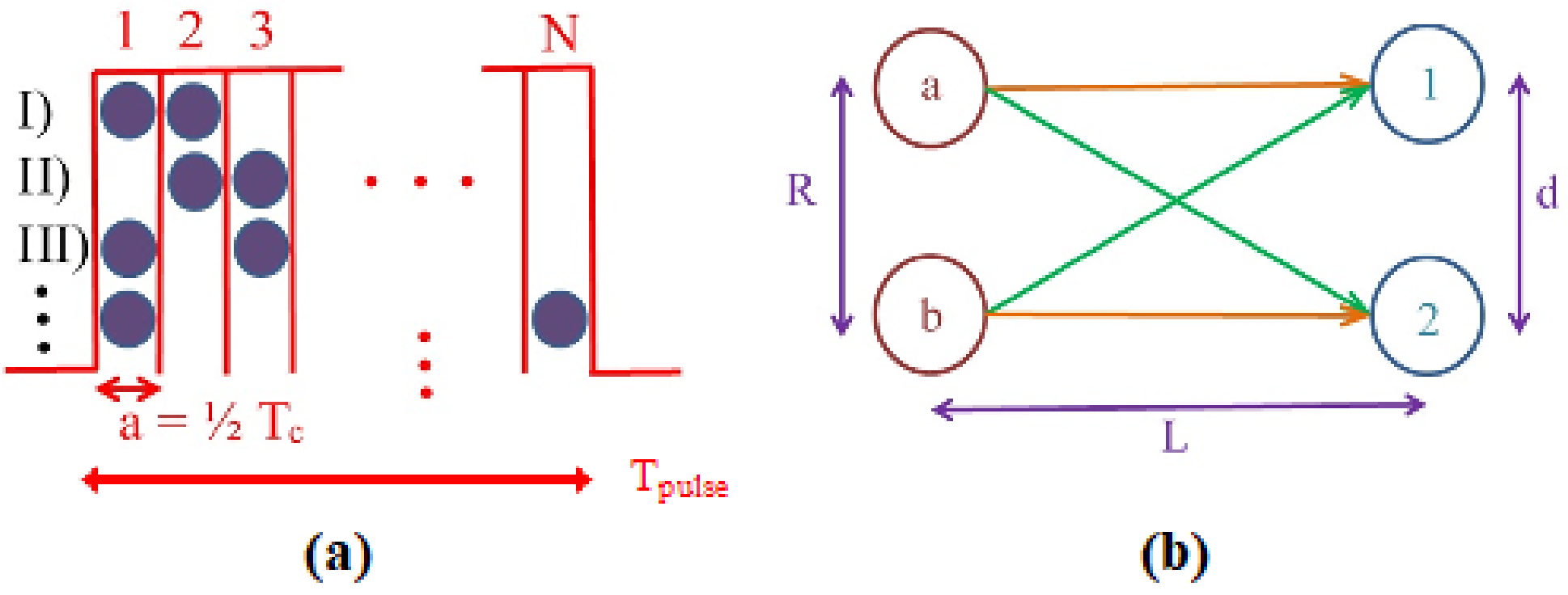

**Figure 1.** Schematics of a pulsed two-electron source. **(a)** The pulse duration $T_{pulse}$ is divided into $N$ equal time segments of duration $a$ and the coherence time $T_c$ extends over two such segments. Two electrons separated by a time interval shorter than $T_c$ are in a two-particle pure state corresponding to an antisymmetric total wavefunction. During the fraction of time when the separation time of the electron pair is longer than the coherence time, they contribute incoherently to the final joint detection probability density. A more realistic beam can be treated as a quantum mixture of all these possible configurations. **(b)** The typical HBT effect diagram with two possible combinations of two-particle states applies to coherent contributions of electron pairs. Incoherent electrons act like distinguishable particles whose effects add up probabilistically. Here the virtual source points are separated by a (spatial or temporal) distance $R$, the detection points are separated by $d$, and the source-to-detector distance is $L$ [12].

In our simulations, propagation to a far-field detector at a fixed distance from the source was performed using the path-integral method [21]. A discussion of this procedure is given in the Methods section. The propagation part of the present problem is discussed in the context of diffraction in time (DIT) of matter waves [22]. While in the case of diffraction in space, the sources, and the detector screen are extended objects in three-dimensional position space, in the case of DIT, the point source and the point detector are held fixed at certain positions. An electron pulse with a finite duration is produced at the point source; in other words, the electrons move through a temporal slit and subsequently diffract in time as monitored by the point detector. Our present HBT problem is an extension of this scenario. Two electrons are distributed in $N$ temporal slits as in Figure 1(a). We then look for the joint detection probability density $P(\tau)$ as a function of the two detection times $t_1$ and $t_2$ with $\tau = t_2 - t_1$, rather than as a function of detection positions as in equation (5). For the case I) in Figure 1, where the two electrons are emitted within one coherence time interval, the two-particle coherent density matrix elements are given by

$$\rho_{S,AS}(t_1,t_2;t_1',t_2') = \langle t_1,t_2|\varphi_{S,AS}\rangle\langle\varphi_{S,AS}|t_1',t_2'\rangle = \varphi_{S,AS}(t_1,t_2)\varphi^*{}_{S,AS}(t_1',t_2'), \tag{6}$$

with

$$\varphi_{S,AS}(t_1,t_2) = \frac{1}{\sqrt{2}}\left[\varphi_1(t_1)\varphi_2(t_2) \pm \varphi_1(t_2)\varphi_2(t_1)\right], \tag{7}$$

as the symmetric and antisymmetric coherent wavefunctions. For the case III) in Figure 1(a), where the two electrons are emitted outside the coherence time window, the elements of the incoherent density operator ( $\rho_{incoh} = \sum p_k |\psi_k\rangle\langle\psi_k|$ ) are subsequently given by

$$\begin{aligned}\rho_{incoh}(t_1,t_3;t_1',t_3') = \; & \frac{1}{2}\langle t_1,t_3|\varphi_p\rangle\langle\varphi_p|t_1',t_3'\rangle + \frac{1}{2}\langle t_1,t_3|\varphi_p{}^P\rangle\langle\varphi_p{}^P|t_1',t_3'\rangle \\ & = \frac{1}{2}\varphi_p(t_1,t_3)\varphi_p^*(t_1',t_3') + \frac{1}{2}\varphi_p{}^P(t_1,t_3)\varphi_p^{P*}(t_1',t_3'),\end{aligned} \tag{8}$$

where the product wavefunctions are

$$\begin{cases}\varphi_p(t_1,t_3) = \varphi_1(t_1)\varphi_2(t_3), \\ \varphi_p{}^P(t_1,t_3) = \varphi_1(t_3)\varphi_2(t_1).\end{cases} \tag{9}$$

No coherence terms are present for this part of the density matrix. From the total partially coherent density operator $\rho$, the final time-dependent joint detection probability density is thus formally written as

$$\begin{aligned}P(t_1,t_2) &= Tr\left(\rho|t_1,t_2\rangle\langle t_1,t_2|\right) = \int\langle t_1',t_2'|\rho|t_1,t_2\rangle\langle t_1,t_2|t_1',t_2'\rangle dt_1'dt_2' \\ &= \int\rho(t_1',t_2';t_1,t_2)\delta(t_1-t_1')\delta(t_2-t_2')dt_1'dt_2' = \rho(t_1,t_2;t_1,t_2).\end{aligned} \tag{10}$$

The symmetric $P_S(\tau)$ and antisymmetric $P_{AS}(\tau)$ components of the normalized joint detection probability density $P(\tau)$ for various pulse durations are shown in Figure 2(a). Both of these functions are made up of an equal mixture of coherent contributions $P_{coh}(\tau)$ associated with the equation (6), and incoherent terms $P_{incoh}(\tau)$ corresponding to the equation (8). The difference is that while the coherent part of $P_S(\tau)$ is symmetric, that of $P_{AS}(\tau)$ is antisymmetric. We thus denote the coherent contribution to $P_S(\tau)$ with $P_{coh,S}(\tau)$ and the coherent contribution to $P_{AS}(\tau)$ with $P_{coh,AS}(\tau)$. Noting to the fact that for the $N$ intervals shown in Figure 1(a) where

$$N = \frac{2T_{pulse}}{T_c}, \tag{11}$$

there are a total of $(N-1)$ coherent contributions and $\frac{1}{2}(N-2)(N-1)$ incoherent ones, the functions $P_{S/AS}(\tau)$ are expressed as

$$P_{S/AS}(\tau) = \frac{2}{N} P_{coh,S/AS}(\tau) + \frac{N-2}{N} P_{incoh}(\tau). \tag{12}$$

Figure 2(a) shows these symmetric (dashed lines) and antisymmetric (solid lines) normalized joint detection probability densities for several different values of pulse duration $T_{pulse}$ for a coherence time of $T_c = 10\,fs$ and a fixed distance between the source and the detector given by $D = 5cm$, all being consistent with the experimentally achievable values. Also, the time of perfectly simultaneous arrival to the detector for the two electrons, known as the anticoincidence time in the context of the fermionic HBT effect, is $T = 50ns$. While $P_{coh,S}(\tau)$ couples with the spin-singlet probability density of its corresponding two electrons, $P_{coh,AS}(\tau)$ goes with the spin-triplet probability density. Clearly, it is the spatially antisymmetric contributions corresponding to the spin-parallel states which give rise to the HBT dips. Now, noting to the fact that for a completely spin-polarized source for which the electron pair take a symmetric spin state, the spatial part of the coherent contribution to the total density matrix is antisymmetric, we conclude that the antisymmetric curves $P_{coh,AS}(\tau)$ in Figure 2(a) are indeed the final joint detection probability densities for fully spin-polarized sources with different degrees of coherence. On the other hand, an unpolarized electron source is represented by a quantum mixture of spin-singlet and spin-triplet states [10]. Therefore, the total joint detection probability density for an unpolarized source is

$$\begin{aligned} P(\tau) &= \frac{2}{N}\left(\frac{1}{4}P_{coh,S} + \frac{3}{4}P_{coh,AS}\right) + \frac{N-2}{N}P_{incoh} \\ &= \frac{1}{2N}P_{coh,S}(\tau) + \frac{3}{2N}P_{coh,AS}(\tau) + \frac{N-2}{N}P_{incoh}(\tau). \end{aligned} \tag{13}$$

Figure 2(b) shows the normalized joint detection probability density $P(\tau)$ for various unpolarized sources using the symmetric and antisymmetric components in Figure 1(a). Figure 3 will subsequently show how the antisymmetric and fully coherent joint detection probability density of Figure 2(a) (the solid blue curve) varies with increasing the electron kinetic energy.

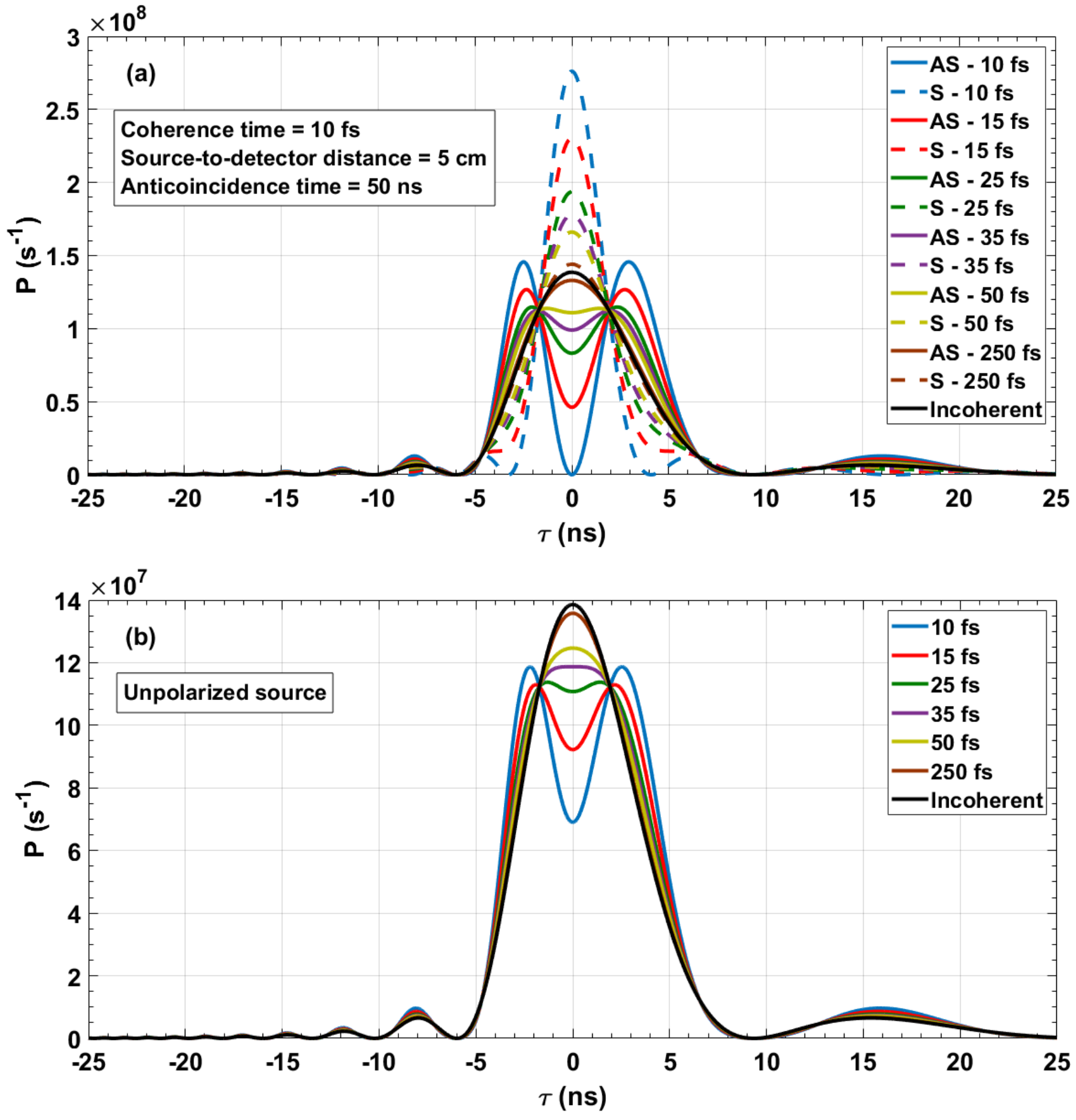

**Figure 2.** Simulation results of the HBT effect for ultra-short pulsed electron beams with arbitrary degrees of coherence showing the two-electron joint detection probability density as a function of the mutual detection time delay. **(a)** The symmetric $S$ and antisymmetric $AS$ components of the normalized joint electron detection probability density for several different pulse durations are shown separately. The assumed experimental condition based on realistic values are denoted in the text box. The antisymmetric curves are also the final joint probability distributions when the pulsed source is completely spin-polarized. The incoherent contribution in the far field is also shown here. **(b)** For an unpolarized source the joint probability density is given by a quantum mixture of the symmetric and antisymmetric components in accordance with equation (13) in the main text. In general, some other interesting features like the flat-top probability distribution for the pulse duration of $35\,fs$ in the present case, due to which mutual detections within an absolute delay time of roughly $1.5ns$ are all equally probable, and the occurrence of local maxima at certain non-zero delay times also deserve future scrutiny. For the latter, the small difference in the absolute values of the delay times corresponding to each pair of maxima is in principle capable of manifesting a kind of two-particle-interference beating in the time domain.

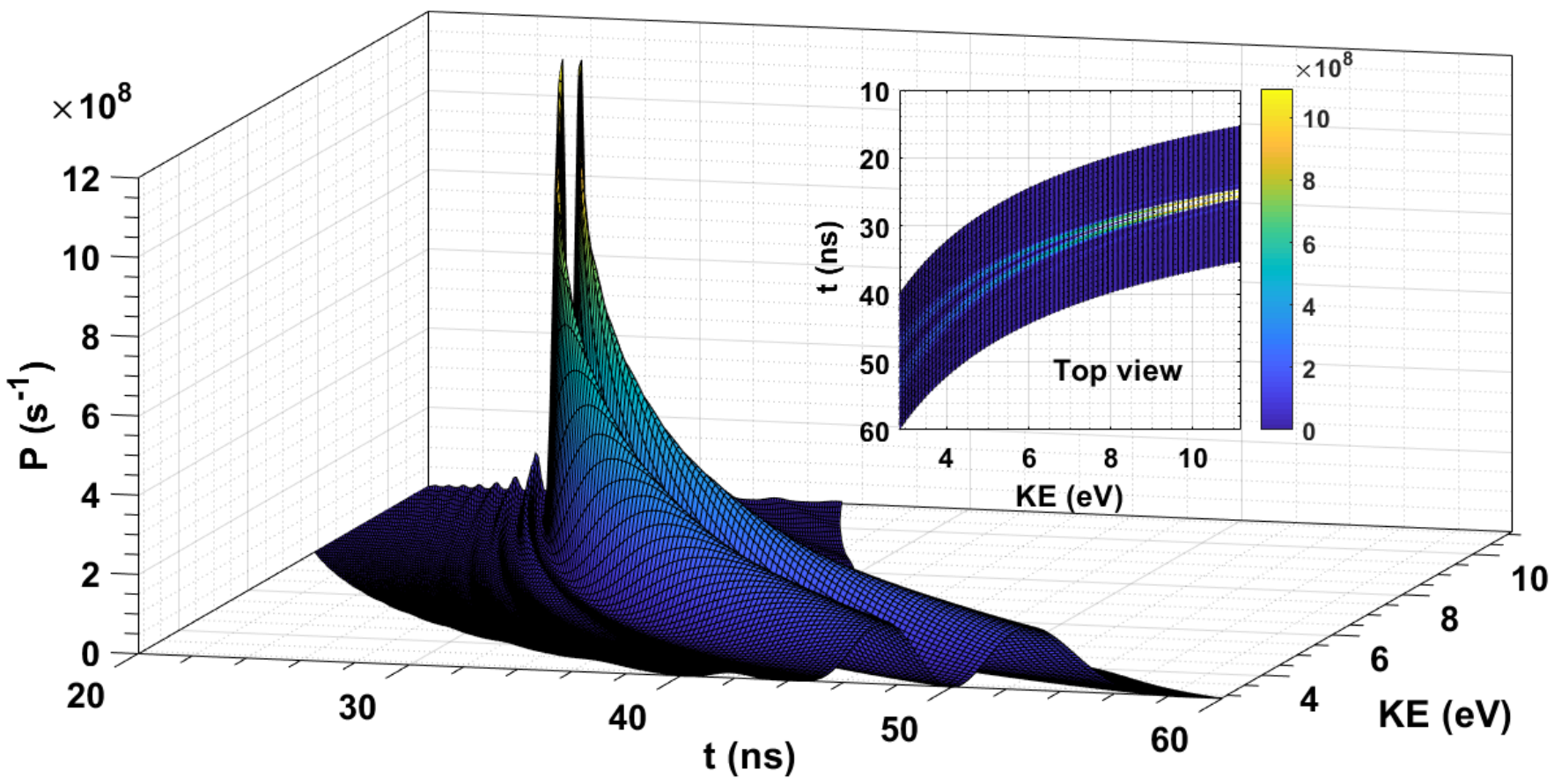

**Figure 3.** Variation of the degenerate polarized two-electron joint probability density with the average electron kinetic energy at practically low energies. The physical parameters are the same as in Figure 2, and the kinetic energy is given by $\frac{m_e}{2}(D/T)^2$ where $D$ is the fixed source-to-detector distance and $T$, the anticoincidence time, is identically the average flight time of the electrons. The front segment with $T = 50ns$ and the lowest kinetic energy of $2.8eV$ shown here overlaps the solid blue line in Figure 2(a). The top view is shown in the inset. As can be noted easily, the HBT dip becomes narrower for progressively increasing values of the kinetic energy (for a fixed distance $D$ ).

At this time, it is useful to define degeneracy and contrast to facilitate further discussion of the results in the above Figures. Since we are interested in the competition of Coulomb pressure and degeneracy pressure, in our electron beams, we need a way of quantifying degeneracy. This is not in the usual textbook sense of counting states of equal energy, but more analogous to statistical mechanics where the degeneracy of a Fermi system is thought of in terms of the thermal wavelength compared to the size of the system. If the thermal wavelength is small compared to the system size, then the Fermi nature of the particles does not come into play. In our systems, we do not have a sense of temperature, so we will quantify how much of the available quantum space is occupied by the electrons. First, consider two simple examples: the ground states of hydrogen and helium atoms, and an electron gas at zero temperature. In the case of the atoms, they have one or two electrons in the 1*s* shell. The degeneracy would be one-half or one in these cases, because either half or all of the available quantum space is occupied. Of course, these atoms have excited orbitals, but these do not contribute to the degeneracy because the electrons have no access to them. In the case of the electron gas, we have a continuous set of momenta which are occupied up to a certain Fermi momentum, and then have no more particles with higher momenta. So the degeneracy above the Fermi momentum would be zero, and below would be one, and, like in the atomic case, the higher levels, though present, do not factor into the degeneracy calculation because the particles have no access to them. From these examples, it is clear that we need the number of occupied states and the total available states, and we need these for a pulse of electrons moving in free space.

The degeneracy $\delta$ is defined as the number of electrons in the coherence volume [5]. The product of the number of emitted electrons $n$ per volume $V$ and the coherence volume $V_c$ gives the number of electrons in the coherence volume as $\delta \approx \left( nV_c / 2V \right)$, where the factor $2$ accounts for the two spin states (this factor would be 2*S*+1 for a general fermion). The coherence volume reflects the space over which the electrons can interfere with each other. In analogy with the electron Fermi gas example above, the unoccupied states above the Fermi surface would be outside the coherence volume. The coherence volume can be obtained via a model (electrons as Gaussian wave packets) or through the energy-time uncertainly principle. The physical volume is determined by the experiment, for example a cylinder with a radius related to the diameter of the aperture emitting the electrons and length related to the electron velocity and pulse time.

Assuming cylindrical symmetry (which is appropriate for nanotip electron sources) the number of electrons in the coherence volume can be written as $\delta \approx \left( nA_c l_c / 2Al \right)$, where $A$ is the cross-section area and $l$ is the volume length. For nanotips, complete transverse coherence $A = A_c$ can be reached and up to ten electrons per laser pulse have been observed [23]. For most of the current manuscript, we limit ourselves to the regime of two electrons per pulse, namely $n = 2$. The coherence volume in the present case is therefore set by $l_c / l = T_c / T_{pulse}$, leading to

$$\delta \approx n \frac{T_c}{2T_{pulse}}. \tag{14}$$

This equation thus links the amount of coherence $T_c / T_{pulse}$ to the degeneracy. To produce the results shown in Figure 2, the coherence time $T_c$ is taken to be $10\,fs$, and the pulse duration $T_{pulse}$ is varied around 50 fs, both consistent with experimentally achieved values. The lack of complete coherence for $T_{pulse} > T_c$, along with the symmetric wavefunction components for an unpolarized electron source, reduces the contrast as seen in Figure 2. In general, arbitrary degrees of coherence and polarization can be treated on the same grounds, and according to our results, both a higher degree of quantum coherence and a higher degree of spin polarization in a beam of two-electron pulses enhance the fermionic HBT antibunching effect.

The HBT contrast can be defined as

$$C_{HBT} = \frac{P_{incoh}^0 - P^0}{P_{incoh}^0 + P^0}, \tag{15}$$

in which the superscript zero indicates that the functions $P_{incoh}\left( \tau \right)$ and $P\left( \tau \right)$ are evaluated at $\tau = 0$ which coincides with the anticoincidence time $T$ defined earlier. The contrast defined this way equals unity when the beam is fully coherent ($T_{pulse} = T_c$) and fully spin-polarized. The contrast is zero in the absence of coherence. Substituting equation (13) into equation (15), the HBT contrast for an unpolarized source is therefore derived as

$$C_{HBT}^{unpol} = \frac{P_{incoh}^{0} - \frac{1}{4}P_{coh,S}^{0}}{\left(N-1\right)P_{incoh}^{0} + \frac{1}{4}P_{coh,S}^{0}} = \frac{1}{2N-1}, \tag{16}$$

where $N$ is given in equation (11) in terms of the characteristic pulse parameters. In the last step use has been made of equation (5) in the time-domain which gives $P_{coh,S}^{0} = 2P_{incoh}^{0}$. For completely polarized two-electron emission there is no symmetric component which subsequently increases the HBT contrast to

$$C_{HBT}^{pol} = \frac{1}{N-1}. \tag{17}$$

As an example, consider the special case of a degenerate beam with $T_{pulse} = T_c$. In that case $C_{HBT}^{pol} = 1$ as expected using equations (11) and (17), whereas for an unpolarized source we get $C_{HBT}^{unpol} = \frac{1}{3}$ using equation (18) instead. This shows a reduction in the contrast by a factor of 3 compared with a similar but polarized beam and identifies a measurable control for the fermionic HBT experiments using controllable ultrafast spin-polarized pulsed electron sources [24]. For $T_{pulse} \gg T_c$ (or equivalently $N \to \infty$), which corresponds to long pulse durations approaching the continuous beam limit, the HBT contrast is small and often negligible even though the spin polarization can still make a difference, noting that

$$\lim_{N\to\infty} \frac{C_{HBT}^{pol}}{C_{HBT}^{unpol}} = \lim_{N\to\infty} \frac{2N-1}{N-1} = 2. \tag{18}$$

It is also instructive to reflect on the absolute amount of reduction in the coincidence count rate at $\tau = 0$ relative to the incoherent illumination of the detectors since this is what is directly sought for in practice [1]. Starting from equation (13), the reduction in the joint detection probability density at the anticoincidence time is obtained as

$$\Delta P^{unpol} = \frac{2}{N}P_{incoh}^{0} - \frac{1}{2N}P_{coh,S}^{0} = \frac{T_c}{T_{pulse}}\left[P_{incoh}^{0} - \frac{1}{4}P_{coh,S}^{0}\right] = \frac{T_c}{2T_{pulse}}P_{incoh}^{0}, \tag{19}$$

for an unpolarized beam, and

$$\Delta P^{pol} = \frac{T_c}{T_{pulse}}P_{incoh}^{0} = 2\Delta P^{unpol}, \tag{20}$$

for a fully polarized beam. This is yet another indication why pulsed electron packets with short durations are significantly more efficient in realizing the HBT effect specifically when compared with continuous beams of free electrons. Now take, for instance, $\Delta P^{unpol} = 1.390 \times 10^{7}$ from the simulation results that led to Figure 2 for a typical pulse duration of $50\,fs$ and assume a repetition rate of $f = 80MHz$. For the sake of comparison with reference [1] let us assume a coincidence time window of $t_w = 26ps$ and a data acquisition time of $T_{acq} = 30h \sim 10^{5}s$. As can be seen in Figure 2, we can safely assume that the joint detection probability density is constant over such a short window at the center. Under the condition that each pulse carries exactly two electrons, the reduced counts in one second, i.e. the reduced count rate denoted by $\Delta R^{unpol}$, is then estimated as

$$\Delta R^{unpol} = \Delta P^{unpol} \times t_w \times f = 2.89 \times 10^{4} cps, \tag{21}$$

which for the above acquisition time gives a total of $\Delta R^{unpol} \times T_{acq} \sim 3 \times 10^{9}$ counts. This is more than a million times bigger than the reported value of $10^{3}$ for the reduced counts in a continuous beam with record-high degeneracy [1]. Such notable enhancements in our signal with ultrashort pulsed electrons will enable us to have a greatly improved control of the experimental conditions in order to segregate the effects of Coulomb repulsion and electron spin. The HBT contrasts corresponding to the joint detection probability distributions shown in Figure 2 are collected and shown in Table 1.

| **$T_{pulse}$ (fs)** | **$^{pol}C_{HBT}$** | **$^{unpol}C_{HBT}$** | **$^{pol}C_{HBT}/^{unpol}C_{HBT}$** | **$\Delta R_{pol}$ (kcps)** | **$\Delta R_{unpol}$ (kcps)** |
|---|---|---|---|---|---|
| **10** | 1.0000 | 0.3350 | 2.985 | 138.6 | 69.5 |
| **15** | 0.5000 | 0.2009 | 2.488 | 92.2 | 46.3 |
| **25** | 0.2500 | 0.1116 | 2.240 | 54.5 | 27.8 |
| **35** | 0.1667 | 0.0772 | 2.159 | 39.6 | 19.9 |
| **50** | 0.1111 | 0.0528 | 2.104 | 27.7 | 13.9 |
| **250** | 0.0204 | 0.0101 | 2.019 | 5.6 | 2.8 |

**Table 1.** HBT contrasts corresponding to the simulation results in Figure 2, where the coherence time is $10\,fs$. The marginal deviation of $C_{HBT}^{unpol}$ from $1/3$ for a degenerate pulse as discussed in the main text is due to computational errors and physical approximations, e.g. the far-field limit, to be addressed in the Methods section. The forth column shows that spin polarization gradually becomes less significant in improving the contrast as the pulse duration grows towards the continuous beam regime. The ratio approaches the value derived in equation (18) for long pulses. In the last two columns, the estimated reduced count rates for a pulsed electron beam repetition rate of $80MHz$ are given in a short detection window of $100\,ps$ for both polarized and unpolarized beams. Refer to the main text for more details. The reduction by a factor of $2$ between the last two columns is of course expected noting that $3/4 - 1/4 = 1/2$.

The possibility of getting single-electron pulses which weaken the signal can also be taken into account on a similar footing. Consider a number $\eta$ of exclusively two-electron pulses produced per second ( $\eta \le f$ ). The reduced count rate is thence given by $\eta \times \left( P_{incoh}^{0} - P^{0} \right)$ times a short coincidence window as before. Now consider a situation where the experimental data point to an average of $\eta_1$ single-electron and $\eta_2$ two-electron pulses emitted per second. Clearly, the reduced count rate in this case is determined by

$$\Delta R = \left( \eta_2 - \eta_1 \right) P_{incoh}^{0} - \eta_2 P^{0}. \tag{22}$$

We thus make a further observation that since for the HBT effect to occur $\Delta R$ must be positive, the following criterion should hold true

$$\frac{\eta_2 - \eta_1}{\eta_2} > \frac{P^{0}}{P_{incoh}^{0}}, \tag{23}$$

for electron antibunching signal to be able to show up. There must be more two-electron pulses in the beam than single-electron ones to observe the HBT effect, for sure, as the necessary condition. It will then suffice to satisfy the inequality in equation (23).

## II-A) Multielectron pulses

We can generalize our model to include pulses with more than two electrons in them. Here we keep the notation $T_{pulse}$ for the shortest time interval which includes two electrons on average and denote the actual electron pulse duration with $\Delta T_e$. In general, the maximum number of electrons in such a pulse is given by

$$\eta_{\max} = 2 \frac{\Delta T_e}{T_c}, \tag{24}$$

which satisfies $T_{pulse} = T_c$. The restriction is of course due to the PEP. For $2 \le \eta \le \eta_{\max}$,

$$T_{pulse} = \left( \frac{\eta_{\max}}{\eta} \right) T_c = \frac{2 \Delta T_e}{\eta}, \tag{25}$$

where equation (24) is used in the last step. This modifies the expression for the number of intervals $N$ in equation (11) as

$$N = \frac{2 T_{pulse}}{T_c} = \frac{4 \Delta T_e}{\eta T_c}. \tag{26}$$

Assuming a Poissonian probability distribution for the number of electrons in each pulse,

$$P_{Poisson}(\eta) = \frac{\langle \eta \rangle^{\eta}}{\eta!} e^{-\langle \eta \rangle}, \tag{27}$$

where

$$\langle \eta \rangle = \frac{\langle R_e \rangle}{f}, \tag{28}$$

the average reduction in the joint probability density at the anticoincidence time, using equations (19) and (25), is obtained as

$$\langle \Delta P^{unpol} \rangle = \sum_{\eta=2}^{\eta_{max}} \left[ \frac{\eta}{\eta_{max}} \times \frac{P_{incoh}^{0}}{2} \times P_{Poisson}(\eta) \right]. \tag{29}$$

$R_e$ in equation (28) is the electron emission rate.

## II-B) Derivation of a partially coherent state based on quantum decoherence theory

So far, we treated the electron beam as a maximally mixed state of coherent and incoherent contributions using the experimentally measurable finite coherence time of electrons in a pulse. Here we discuss an analytical method to obtain a partially coherent two-electron state starting from a pure entangled state with well-defined symmetries for the two electrons on the source as required by the PEP. To this goal, we model the emission process based on a two-state atom with a large electron-pair emission cross-section. The emitted electrons are assumed to initially be in an entangled state together with their emitter. Once the electrons leave the nanotip, the emitter states are traced out from the original pure three-particle entangled state leaving behind a two-electron partially coherent state. This way we treat the emitter as the "environment" for the electron-pair system in the language of the quantum decoherence theory [25,26]. The lack of complete coherence in the beam can thereby be systematically attributed to the initial entanglement of the electron pair with their two-state emitter during the emission time interval $\Delta t \sim \hbar/\Delta E$, with $\Delta E = E_e - E_g$ being the energy gap between the excited and the ground states of the emitter involved in the process. In the following, we first introduce a suitable entangled three-particle pure state with density operator

$$\rho_{12E} = |\psi_{12E}\rangle\langle\psi_{12E}|, \tag{30}$$

where the subscripts denote the two electrons and the environmental particle $E$. We can then move on and compute the decohered two-electron state by tracing out the environmental states:

$$\rho_{12} = Tr_E \rho_{12E}. \tag{31}$$

To obtain the spatial state involved in the HBT effect, we also average out the two-particle spin states $\{|\uparrow\downarrow\rangle, |\downarrow\uparrow\rangle\}$ by taking the trace over them:

$$\overline{\rho}_{12} = Tr_s \rho_{12}. \tag{32}$$

Optionally, one may also look into the single-electron density operator by tracing out one of the electron states:

$$\rho_1 = Tr_2 \overline{\rho}_{12}. \tag{33}$$

The last two steps in equations (29) and (30) are interchangeable. It is important to note that it is the two-electron state which is to be propagated towards the detector in simulations of the HBT effect.

In what follows, the electron spin-up (-down) is denoted with $\uparrow$ ($\downarrow$). $|t_a\rangle$, $|t_b\rangle$, and $|t_c\rangle$ are three single-electron orbitals at the source corresponding to $N = 3$ intervals (see Figure 1(a)). Unless the tensor-product sign $\otimes$ is explicitly used, factorizable states like $|t_a\rangle \otimes |t_b\rangle$ are simply written as $|t_a t_b\rangle$. In all such cases, the particle exchange number is 1 (2) for the left (right) eigenvalue. This must be kept in mind in inspecting the symmetry of the composite states. Note specifically that $|\psi_{12E}\rangle$ must be antisymmetric under the exchange of the two electrons. $|g_E\rangle$ and $|e_E\rangle$ indicate the ground and excited states of the emitter, that is the environmental particle, respectively.

Let us consider now the following normalized entangled state:

$$|\psi_{12E}\rangle = \frac{1}{4}\left\{\begin{array}{l} \left[\left(|t_a t_b\rangle + |t_b t_a\rangle\right) \otimes \left(|\uparrow\downarrow\rangle - |\downarrow\uparrow\rangle\right) \otimes |g_E\rangle\right] + \\ \left[\left(|t_b t_c\rangle + |t_c t_b\rangle\right) \otimes \left(|\uparrow\downarrow\rangle - |\downarrow\uparrow\rangle\right) \otimes |g_E\rangle\right] + \\ \left[\left(|t_a t_c\rangle + |t_c t_a\rangle\right) \otimes \left(|\uparrow\downarrow\rangle - |\downarrow\uparrow\rangle\right) \otimes |g_E\rangle\right] + \\ \left[\left(|t_a t_c\rangle - |t_c t_a\rangle\right) \otimes \left(|\uparrow\downarrow\rangle + |\downarrow\uparrow\rangle\right) \otimes |e_E\rangle\right] \end{array}\right\}, \tag{34}$$

which presents a case where the two electrons are in a spatially symmetric state and the singlet spin state before emission and are coherent at all times. The quantum correlation modeled here is such that excitation of the emitting atom to the energy state $|e_E\rangle$ by an incident laser pulse is accompanied by the two electrons taking an antisymmetric spatial component and a symmetric spin part. We represent $\rho_{12}$ given by equation (31) on the basis states

$$\left\{ \begin{array}{l} |t_a t_b;\uparrow\downarrow\rangle, |t_a t_b;\downarrow\uparrow\rangle, |t_b t_a;\uparrow\downarrow\rangle, |t_b t_a;\downarrow\uparrow\rangle, |t_a t_c;\uparrow\downarrow\rangle, |t_a t_c;\downarrow\uparrow\rangle, \\ |t_c t_a;\uparrow\downarrow\rangle, |t_c t_a;\downarrow\uparrow\rangle, |t_b t_c;\uparrow\downarrow\rangle, |t_b t_c;\downarrow\uparrow\rangle, |t_c t_b;\uparrow\downarrow\rangle, |t_c t_b;\downarrow\uparrow\rangle \end{array} \right\}, \tag{35}$$

in order from left to right (top to bottom) for each row (column) of the matrix. Consequently,

$$\rho_{12} \doteq \frac{1}{16}\begin{pmatrix} 1 & -1 & 1 & -1 & 1 & -1 & 1 & -1 & 1 & -1 & 1 & -1 \\ -1 & 1 & -1 & 1 & -1 & 1 & -1 & 1 & -1 & 1 & -1 & 1 \\ 1 & -1 & 1 & -1 & 1 & -1 & 1 & -1 & 1 & -1 & 1 & -1 \\ -1 & 1 & -1 & 1 & -1 & 1 & -1 & 1 & -1 & 1 & -1 & 1 \\ 1 & -1 & 1 & -1 & 2 & 0 & 0 & -2 & 1 & -1 & 1 & -1 \\ -1 & 1 & -1 & 1 & 0 & 2 & -2 & 0 & -1 & 1 & -1 & 1 \\ 1 & -1 & 1 & -1 & 0 & -2 & 2 & 0 & 1 & -1 & 1 & -1 \\ -1 & 1 & -1 & 1 & -2 & 0 & 0 & 2 & -1 & 1 & -1 & 1 \\ 1 & -1 & 1 & -1 & 1 & -1 & 1 & -1 & 1 & -1 & 1 & -1 \\ -1 & 1 & -1 & 1 & -1 & 1 & -1 & 1 & -1 & 1 & -1 & 1 \\ 1 & -1 & 1 & -1 & 1 & -1 & 1 & -1 & 1 & -1 & 1 & -1 \\ -1 & 1 & -1 & 1 & -1 & 1 & -1 & 1 & -1 & 1 & -1 & 1 \end{pmatrix}. \tag{36}$$

The dot in $\doteq$ indicates that the right-hand side is representation-dependent. Using the equation (32), the spin-averaged two-electron density matrix represented on the set of basis states given by

$$\left\{ |t_a t_b\rangle, |t_b t_a\rangle, |t_a t_c\rangle, |t_c t_a\rangle, |t_b t_c\rangle, |t_c t_b\rangle \right\}, \tag{37}$$

is computed as

$$\bar{\rho}_{12} \doteq \frac{1}{8}\begin{pmatrix} 1 & 1 & 1 & 1 & 1 & 1 \\ 1 & 1 & 1 & 1 & 1 & 1 \\ 1 & 1 & 2 & 0 & 1 & 1 \\ 1 & 1 & 0 & 2 & 1 & 1 \\ 1 & 1 & 1 & 1 & 1 & 1 \\ 1 & 1 & 1 & 1 & 1 & 1 \end{pmatrix}. \tag{38}$$

Clearly, the sub-space $\left\{ |t_a t_c\rangle, |t_c t_a\rangle \right\}$ has lost its coherence in this process while both $\left\{ |t_a t_b\rangle, |t_b t_a\rangle \right\}$ and $\left\{ |t_b t_c\rangle, |t_c t_b\rangle \right\}$ sub-spaces are remained coherent which is what we desired for. The generalization to longer pulses with $N > 3$ is straightforward.

## III. Conclusions

Our main goal in this article was to demonstrate that ultra-short pulsed electron beams can enhance the Hanbury Brown-Twiss (HBT) antibunching signal, and that the enhancement will be even larger for ultrafast spin-polarized electron sources. Through a general description of the realistic partially coherent pulsed electron beams in terms of a maximal quantum mixture of coherent and incoherent contributions, we showed that this is indeed the case, comparing ultrashort electron pulses with the limit of continuous free electron beams for which few experimental reports along with their critical analyses exist in the literature. Greatly enhanced signal-to-noise-ratios as anticipated from our simulation results will enable us to have an improved control over the experimental parameters which is specifically helpful in segregating the effects of electron spin and Coulomb repulsion between free electron pairs in our experiment with pulsed electrons to be reported in the near future. As discussed in the literature, the latter can lead to antibunching signals which are rooted in coulombic interactions rather than spin statistics. Therefore, for a decisive observation of the Pauli exclusion principle (PEP) for free electrons, implementation of ultrashort pulsed electron beams with durations on the order of the coherence time of the electrons in each pulse seem to be compulsory according to the results given in the present article. Our analysis also showed that in this regime, a polarized pulsed electron source can further enhance the antibunching signal by up to a factor of $3$ for degenerate pulses with duration identical to the coherence time of the two electrons carried by the pulse packet. We derived expressions for an arbitrarily defined HBT contrast, the absolute reduced count rate, and a criterion for the HBT effect to happen where single-electron pulses are also present and showed how the model can be generalized to take into account multi-electron pulses. Finally, we provided a generic method that enabled us to treat the concept of quantum "partial" coherence on a more rigorous ground based on ideas borrowed from quantum decoherence theory. We believe the procedure offered here has the potential to address the lack of complete coherence in free electron beams based on emission mechanisms originated in an entangled state of the emitters and the to-be-emitted electrons by bridging to the quantum decoherence theory. We hope that our treatment of the problem can spin a new twist on observations of the electronic HBT effect as well as the quantum interference of identical particle pairs, in general.

## IV. Methods and further discussion

The problem discussed in this article contained two main parts: 1) Writing down an appropriate initial two-electron state, and 2) Propagating the state to a detector located at the far-field limit of the emitter. The first step was discussed in detail in the main text. The simulation results after the second state were given in Figure 2 and Table 1, and were subsequently discussed and analyzed in the Results and Discussion section. Here we give an overview of the path-integral (PI) method [21] that we have chosen to use for the propagation problem in the context of matter wave diffraction in time (MWDIT). The latter was previously introduced after Figure 1 and before equation (6). The MWDIT is sketched schematically in Figure 4.

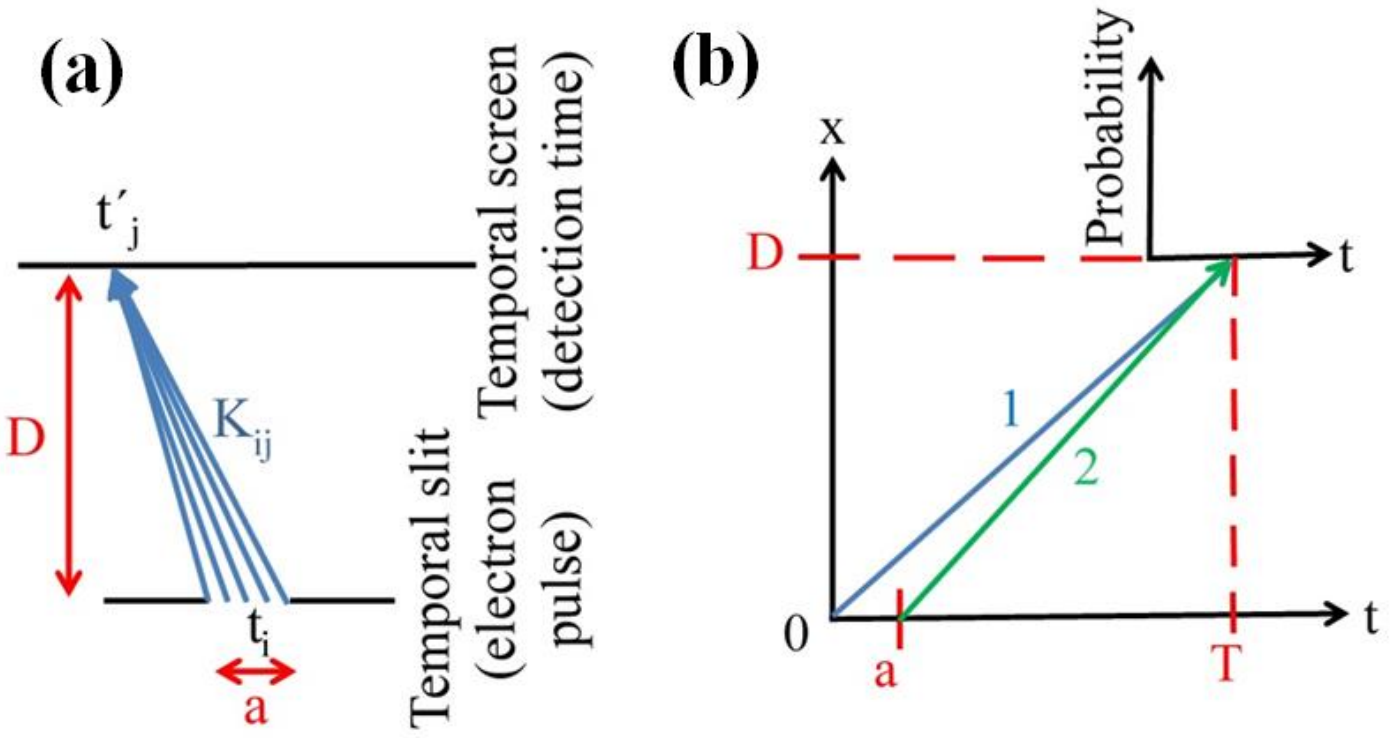

**Figure 4.** Schematics of the MWDIT. **(a)** When shutter is open for a time $a$ electrons are emitted toward a detector held fixed at a sufficiently long distance $D$ from the source (to satisfy the far field approximation) which can register the arrival time of each emitted electron. This is exactly the case for the emission of an electron pulse with duration $a$ that embeds one electron. The observed diffraction pattern which indicates wave packet broadening can conveniently be computed using the PI method where the probability amplitude is obtained by integrating the accumulated complex phase $e^{iS/\hbar}$, with $S$ being the action defined by the temporal integral of the Lagrangian – here for free space – over all possible phase-space trajectories. A few representative paths to a single detection time $t_j$ from a few of the source points $t_i$'s are sketched here. $K_{ij}$ is the Feynman kernel given by the above phase, whose modulus square gives the probability for an electron to go from $t_i$ to $t_j$ along a possible path. **(b)** $x-t$ diagram of MWDIT in one-dimensional space. The main peak of the single-particle diffraction pattern occurs at time $T \gg a$. For this to be the case, same as in light optics, the wavelets must arrive in-phase at the detector at time $T$ through the two paths labeled 1 and 2 originated from the source boundaries.

The free-space PI kernel for path number $k$ is given by

$$K\left(t_k^i;t_k^f\right)=\sqrt{\frac{m_e}{i2\pi\hbar\left(t_k^f-t_k^i\right)}}\exp\left[\frac{im_eD^2}{2\hbar\left(t_k^f-t_k^i\right)}\right], \tag{39}$$

between an initial and a final time denoted by superscripts $i$ and $f$, respectively. $m_e$ is the electron mass and $D$ is the constant distance between the source and the detector. Same as in optics the wavelets propagated through the paths 1 and 2 in Figure 4(b) must arrive at the detector in-phase at time $T$ as the condition for constructive interference. The accumulated phase through each of these two paths is determined by the exponential term in equation (39). Also, for a plane wavefunction at the source, the initial phase is given by $\omega t$ where $\omega$ is the angular frequency. The condition of constructive interference at time $T$ will thus enable us to determine the expression for $\omega$ consistent with this requirement that we can subsequently use in our simulations. The constructive interference condition therefore becomes

$$\xi_2^0 + \xi_2 - \xi_1^0 - \xi_1 = 0, \tag{40}$$

where $\xi_i$ with $i = 1, 2$ is the accumulated phase over each of the two paths. The superscript $0$ indicates the initial phase on the source. Plugging in the correct expression for each phase function we obtain

$$\frac{m_e D^2}{2\hbar T}\left[\frac{1}{1+\frac{a}{T}} - 1\right] - \omega a = 0, \tag{41}$$

as the exact phase equation. In the far-field limit, $a \ll T$ which gives

$$\omega = \frac{-m_e D^2}{2\hbar T^2}. \tag{42}$$

For the sake of comparison, for the more familiar case of diffraction in space, the corresponding wave-number obtained by the same method is

$$\kappa = \frac{m_e \Delta}{\hbar \tau}, \tag{43}$$

where in that case, $\tau$ is the constant detection time and $\Delta$ is the point on the screen where the diffraction peak will show up after a sufficiently large number of single-electron detection events. The far-field approximation that leads to equation (43) can be expressed as $\alpha \ll \Delta$, where $\alpha$ is the spatial width of the source.

It is also instructive to find an expression for the first zeros of the diffraction pattern as a consistency check to make sure that our simulation results comply with it. The condition for destructive interference at a time $t_0 \gg a$ is given by

$$\xi_2^0 + \xi_2 - \xi_1^0 - \xi_1 = \pm 2\pi, \tag{44}$$

as in optics for a wide source. It follows that the times of the first zeros are given by

$$t_0 = \frac{1}{\sqrt{\frac{\pm 4\pi\hbar}{m_e a D^2} + \frac{1}{T^2}}}, \tag{45}$$

where the minus (plus) sign corresponds to the first zero on the leading (trailing) edge of the diffraction pattern.

The double-slit experiment in time can also be treated similarly. Two shutters unblocking an incoming single-electron matter wave is then equivalent to the case where one electron can either be found in the interval $1$, corresponding to slit number $1$, or in time interval $2$, corresponding to slit number $2$, with negligible detection probability elsewhere. When the slits are separated by a time interval, typically on the order of the slit duration, a temporal interference pattern can be observed by measuring the arrival times of electrons at the detector for an adequately large number of detection events. One must keep in mind that the double-slit experiment is a single-particle interference event. For this to be the case for MWDIT in the presence of a physical synchronous double-shutter, the overlap between the emitted electrons must be negligible which is the case for low degeneracies. In contrast, in the electronic HBT effect, the antibunching signal is created by overlapping electrons whose composite state is antisymmetric in compliance with the PEP. As we argued throughout this article, there is a higher chance for this to happen with ultrashort electron pulses as compared with continuous beams of electrons. Where the main constructive interference fringe is observed for the double-slit setup as shown in Figure 5(d), an HBT dip occurs under similar experimental conditions as in Figure 2(a). In accordance with our explanations given in the Results and Discussion section, for electrons outside a coherence volume, the situation is resembled by two independent shutters whose individual diffraction patterns contribute incoherently to the final effect noting that each single shutter spans one coherence volume with duration $\Delta t \sim \hbar/\Delta E$ and is hindered by the PEP to host more than two identical electrons. Two electrons in a single coherence volume then contribute to the HBT dip. In our treatment of free electron pulses discussed in this manuscript, no double-slits cross the beam path. The only effects present are therefore incoherent contributions mimicking single-slit single-electron diffraction pattern and the antibunching signal due to coherent, i.e. quantum mechanically correlated, electron pairs. One may thus envision spatial and temporal double-slit experiments with highly degenerate pulsed electron beams manifesting an interplay between coherent single particle interferences and the two-electron HBT effect. The incoherent electrons in our model will then undergo double-slit interference with maximum peak heights of up to twice the incoherent diffraction pattern at the HBT anticoincidence time which will reduce the antibunching signal even further. From the perspective of the single electron interference, it is the very presence of two-fermion interference that diminishes the first order fringe visibility. The question begging an answer is then whether the double-slit will decohere the coherent electron pairs or not. If it does, no HBT signal will be observed. Any hint of antibunching will then indicate that some coherent pairs of electron are not affected by the double-slit, based on which a “which-way” experiment could follow if it implied crossing of the correlated pair solely through one or the other of the two slits. Another possibility is, of course, the coherent electron pair in each ultra-short pulse acting as a single boson with twice the mass of one electron undergoing a double-slit arrangement and giving rise to single-particle interference patterns while the uncorrelated electrons in each pulse tend to generate single-electron interference patterns which are broader by a factor of two. Figure 7 demonstrates this latter conjecture. Before then, Figure 5 gives an overview of the MWDIT and Figure 6 compares the fermionic HBT effect with that of the double-slit experiment.

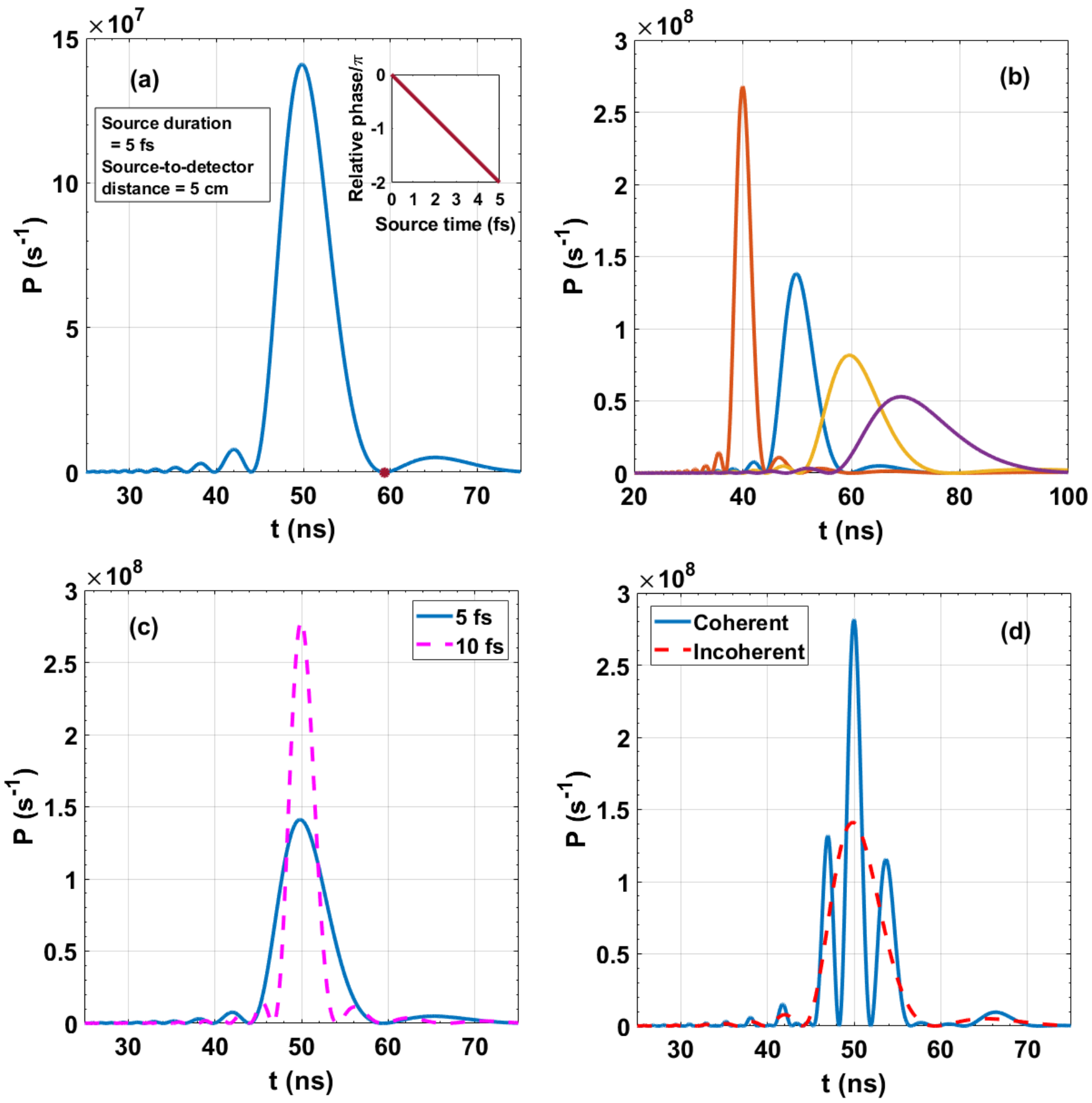

**Figure 5.** An overview of the MWDIT showing single-electron detection probability density as a function of the detection time. **(a)** Single-slit electron DIT analogous to a one-electron pulse broadening. The relevant parameters are given in the text box. The inset shows the phase difference of the trajectories originated from different points on the source at the time of the first zero at the detector position, calculated using equation (45), and indicated by a circle on the leading edge of the diffraction pattern. The total phase difference, which is computed relative to the phase from the trajectory originated at $t=0$, is $-2\pi$ as anticipated. The reader less familiar with interferometry must note that in this case, for every trajectory originated in one half of the source, there is another trajectory started from a point on the second half which arrives at the first zero with an absolute phase difference of $\pi$ relative to the first path, hence a destructive interference. In addition, unlike in the more familiar case of diffraction in space, the MWDIT pattern is clearly non-symmetric in time. The reason is that the electrons on the leading edge take a longer time to travel the same distance $D$ as compared with those on the trailing edge. In other words, they have smaller momenta and consequently longer de Broglie wavelengths consistent with this observation. **(b)** Faster electron packets broaden less than slower ones in traveling the same distance from the source. **(c)** The resultant pattern for two different source durations are compared. Shorter wavepackets broaden faster, familiar effect from non-relativistic quantum mechanics. **(d)** Coherent and incoherent contributions in a double-slit DIT. All parameters are the same as in **(a)** with the slit separation being equal to each slit duration.

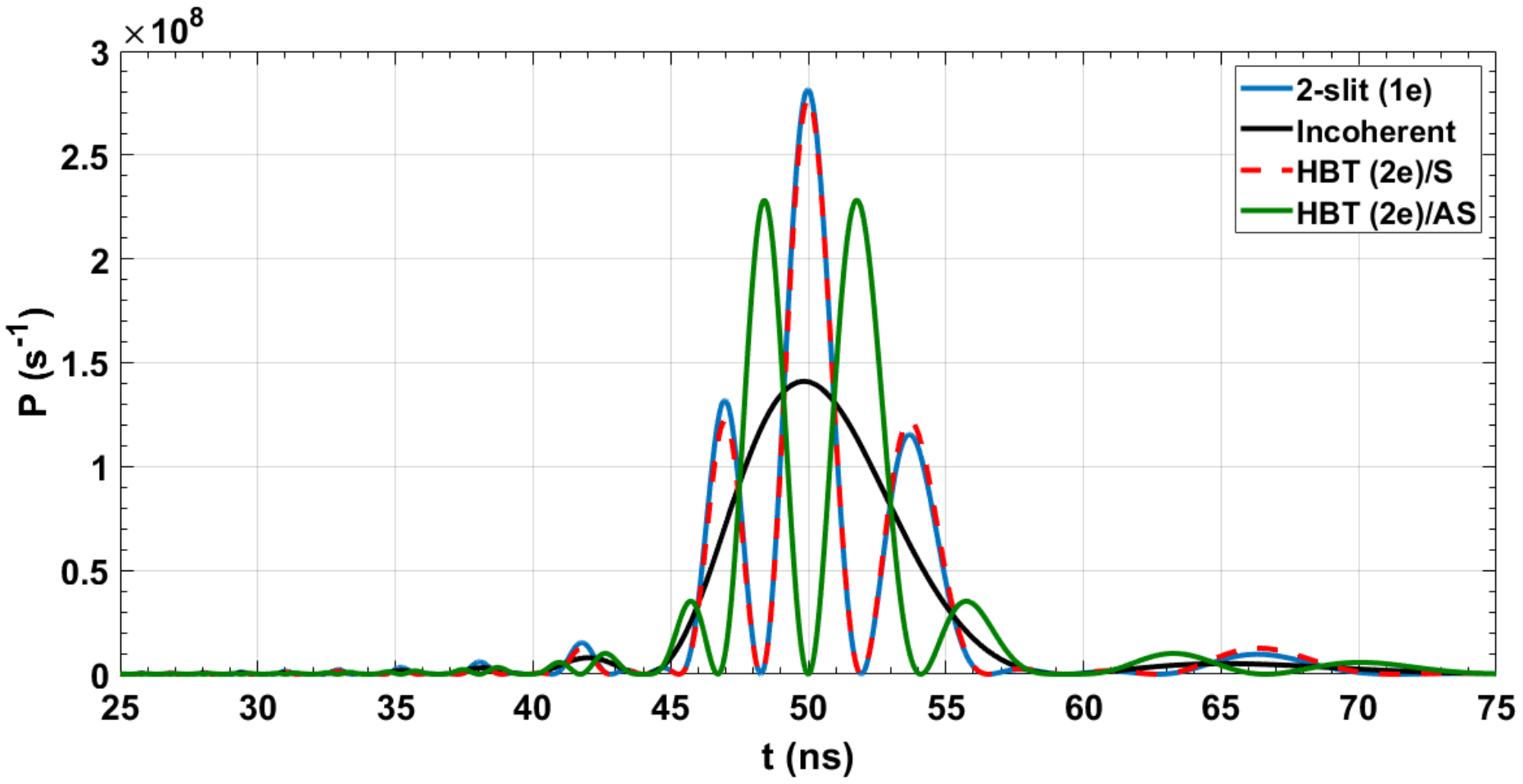

**Figure 6.** Comparison between the single-electron probability density in a double-slit experiment with the symmetric and antisymmetric components of the two-electron joint detection probability density in the fermionic HBT effect for the same experimental conditions which includes an equal separation of the two correlated electrons in the HBT case compared with the slit separation in the 1-electron double-slit scenario. The numerical values of the relevant parameters are identical to those given in Figure 5.

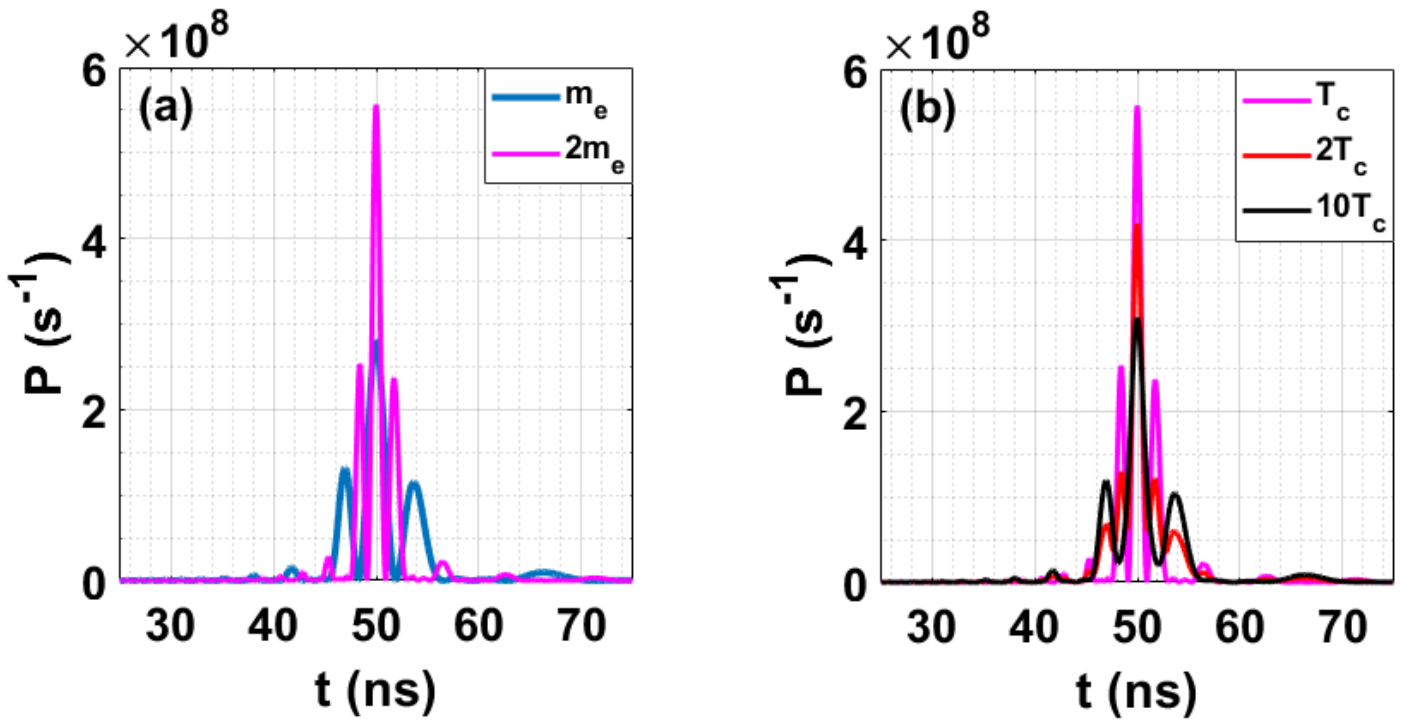

**Figure 7.** One of the conjectures discussed above before Figure 5 is demonstrated here. The physical parameters are identical to those of Figure 5. As elaborated in the main text, the 2-electron pulses propagating in free space will lead to HBT dips with various degrees of contrast depending on their degrees of coherence. Now what happens if a double-slit is placed in the path of the pulsed electron beam? One possibility is that when the two electrons are within one coherence time they act as a boson with twice the mass of one electron while two uncorrelated electrons will each give rise to the single-electron interference pattern. The two contributions are compared in **(a)**. Keeping up with this conjecture and as shown in **(b)** for three different pulse durations, for a completely coherent beam, namely the degenerate electron beam, the bosonic pattern will then be obtained whereas for partially coherent beams the single-electron interference pattern will tend to dominate progressively as the degree of coherence diminishes with increasing the ratio between the pulse duration and the coherence time. Not overviewed here is the other conjecture discussed above according to which the fermionic HBT effect would persist to show up in spite of the presence of a double-slit in the path of the beam.

## V. Error analysis

Consider the case $N = 3$. From our model based on a maximally mixed state we can write

$$P(\tau) = \frac{1}{3} P_{coh}^{1,2}(\tau) + \frac{1}{3} P_{coh}^{2,3}(\tau) + \frac{1}{3} P_{incoh}^{1,3}(\tau), \tag{46}$$

where the superscripts denote the interval number. To save computation time and resources, which is significant for $N \gg 1$, we approximate equation (46) with

$$P(\tau) = \frac{2}{3} P_{coh}^{1,2}(\tau) + \frac{1}{3} P_{incoh}^{1,3}(\tau), \tag{47}$$

noting that in the far-field approximation $P_{coh}^{2,3}(\tau) \approx P_{coh}^{1,2}(\tau)$. Similarly, for $N > 3$ we use the same approximation for the coherent terms. The incoherent terms are also approximated with $P_{incoh}^{1,3}(\tau)$. In all such cases,

$$P_{coh,incoh}^{i,j}(\tau) = P_{coh,incoh}^{1,2 or 3}(\tau) + e_{coh,incoh}^{i,j}, \tag{48}$$

$e_{coh,incoh}^{i,j} = \max\left(e_{coh,incoh}^{i,j}(\tau)\right)$ is the upper limit of the error. Table 2 shows the relative errors in the maximum value of $P_{incoh}(\tau)$ for simulation results in a number of examples using parameters of Figure 2.

| $N_i$ , $N_j$ | Relative error |
|---|---|
| **1,3** | 0 |
| **1,10** | $1.4\times10^{-6}$ |
| **8,10** | $2.1\times10^{-6}$ |
| **1,50** | $7.2\times10^{-6}$ |
| **1,100** | $1.5\times10^{-5}$ |
| **98,100** | $3.0\times10^{-5}$ |

**Table 2.** Examples of the error in the maximum value of $P_{incoh}(\tau)$ relative to the case used as a valid approximate.

Lastly, it is important to note that the HBT contrasts $C_{HBT}^{unpol}$ and $C_{HBT}^{pol}$ given in equations (16) and (17) are clearly unaffected by the computational errors discussed here. In addition, marginal errors when removed by using the exact curves instead do not cause any noticeable change in the plotted curves in practice as long as we stick with the range of applicability of the far-field approximation.

## Acknowledgements

We acknowledge support for this work by the National Science Foundation (NSF) under the award number PHY-1602755.